\documentstyle[11pt,newpasp,twoside,epsfig]{article}
\begin{document}
\title{Constraints on the equation of state of ultra-dense matter from
       observations of neutron stars}
\author{M. H. van Kerkwijk}
\affil{Astronomical Institute, Utrecht University, P.O. Box 80000,
       3508~TA Utrecht, The Netherlands}

\begin{abstract}
I discuss constraints on the equation of state of matter at
supra-nuclear densities that can be derived from observations of
neutron stars.  I focus on recent work on Vela X-1, which may well be
substantially more massive than the canonical $1.4\,M_\odot$, and on
the prospects offered by the `isolated' or `thermally-emitting'
neutron stars.
\end{abstract}

\section{The equation of state for ultra-dense matter}

To understand the core collapse of massive stars, the supernova
phenomenon, and the existence and properties of neutron stars,
requires knowledge of the equation of state (EOS) for matter at
supra-nuclear density.  The EOS is determined by the behaviour of
elementary particles at close proximity to each other and hence is of
fundamental physical interest.  It is modeled using
quantum-chromodynamics calculations, but these are not developed well
enough to determine the densities at which, e.g., meson condensation
and the transition between the hadron and quark-gluon phases occur.
At densities slightly higher than nuclear and at high temperatures,
the model predictions can be compared with the results of heavy-nuclei
collision experiments.  For higher densities and low temperatures,
however, this is not possible; the models can be compared only with
neutron-star parameters.  Recent reviews of our knowledge of the EOS,
and the use of neutron stars for constraining it, are given by
Heiselberg \& Pandharipande (\cite{heisp:00}), Lattimer \& Prakash
(\cite{lattp:00}, \cite{lattp:01}), and Balberg \& Shapiro
(\cite{balbs:00}).

The different models for the EOS predict highly different mass-radius
relations, and a direct constraint on the EOS would be set by a
simultaneous measurement of the radius and mass of a neutron star.
This has not yet been possible, and observational tests have been
limited to predictions for extrema, such as the maximum possible mass
and the minimum possible spin or orbital period.  For instance, for
EOS with a phase transition at high densities, such as Kaon
condensation (Brown and Bethe \cite{browb:94}), only neutron stars
with mass $<\!1.5\,M_\odot$ could exist (for larger masses, a black
hole would be formed).

So far, susceptibility to systematic errors and modeling
uncertainties have befuddled most attempts to constrain the EOS
observationally (e.g., radius determinations from X-ray bursts, Lewin
et al.\ \cite{lewivpt:93}; innermost stable orbit from kHz QPOs, Van
der Klis \cite{vdkl:00}).  The only accurate measurements are the
fastest spin period and some precise masses.  The former, 1.5\,ms,
excludes the stiffest EOS (PSR~B1937+214; Backer et
al. \cite{back&a:82}); the latter I discuss below.

\section{Neutron star masses}

Most mass determinations have come from radio timing studies of
pulsars; see Thorsett \& Chakrabarty \cite{thorc:99} for an excellent
review.  The most accurate ones are for pulsars that are in eccentric,
short-period orbits with other neutron stars, such as the Hulse-Taylor
pulsar PSR B1913+16, in which several non-Keplerian effects on the
orbit can be observed: the advance of periastron, the combined effect
of variations in the second-order Doppler shift and gravitational
redshift, the shape and amplitude of the Shapiro delay curve shown by
the pulse arrival times as the pulsar passes behind its companion, and
the decay of the orbit due to the emission of gravitational waves.
Thorsett \& Chakrabarty found that for all radio-pulsar binaries, the
masses were consistent with being in a surprisingly narrow range,
which can be approximated with a Gaussian distribution with a standard
deviation of only $0.04\,M_\odot$.  The mean of the
distribution is $1.35\,M_\odot$, close to the ``canonical'' value of
$1.4\,M_\odot$.

Neutron-star masses can also be determined for some binaries
containing an accreting X-ray pulsar, from the amplitudes of the X-ray
pulse delay and optical radial-velocity curves in combination with
constraints on the inclination (the latter usually from the duration
of the X-ray eclipse, if present).  This method has been applied to
about half a dozen systems (Joss \& Rappaport \cite{jossr:84}; Nagase
\cite{naga:89}; Van Kerkwijk, Van Paradijs, \& Zuiderwijk
\cite{vkervpz:95}).  The masses are generally not very precise, but
are consistent with $\sim\!1.4\,M_\odot$ in all but one case.

The one exception is the X-ray pulsar Vela X-1, which is in a 9-day
orbit with the B0.5\,Ib supergiant HD~77581.  For this system, a
rather higher mass of around $1.8\,M_\odot$ has consistently been
found ever since the first detailed study in the late seventies (Van
Paradijs et al.\ \cite{vpar&a:77}; Van Kerkwijk et al.\
\cite{vker&a:95}).  A problem with this system, however, is that the
measured radial-velocity orbit, on which the mass determination
relies, shows strong deviations from a pure Keplerian radial-velocity
curve.  These deviations are correlated within one night, but not from
one night to another.  A possible cause could be that the varying
tidal force exerted by the neutron star in its eccentric orbit excites
high-order pulsation modes in the optical star which interfere
constructively for short time intervals.

\begin{figure}[t!]
\centerline{\epsfig{figure=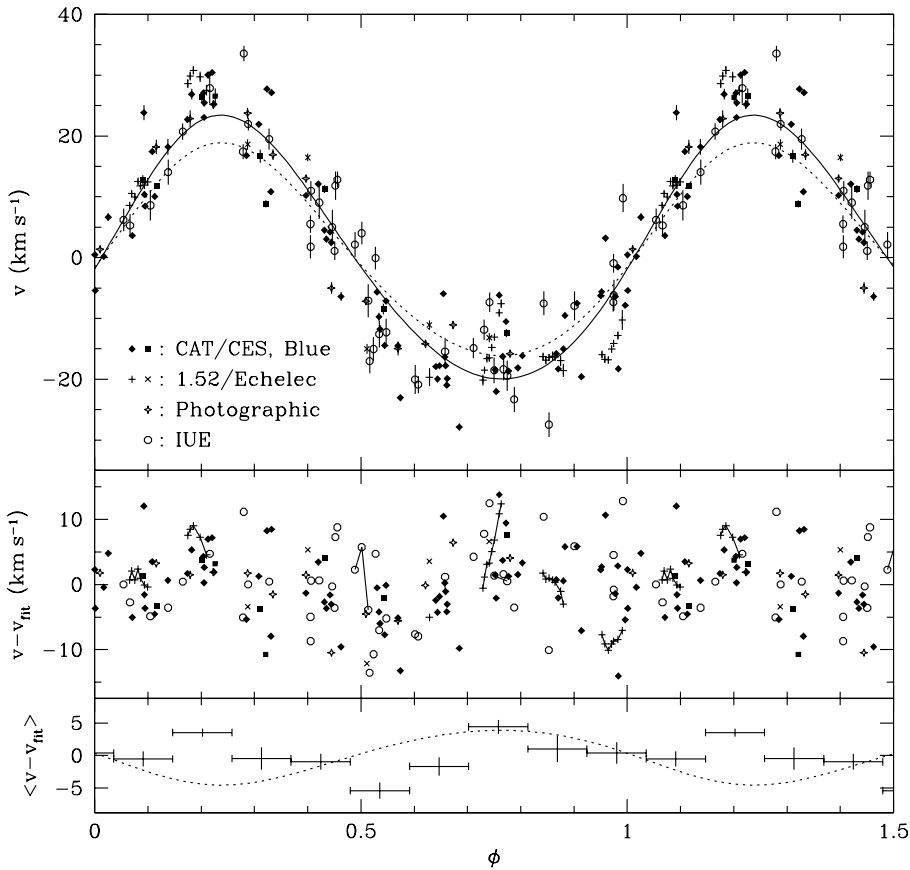,width=0.9\hsize}}
\caption[]{Radial-velocity measurements for HD\,77581, the optical
counterpart to Vela X-1.  Overdrawn is the Keplerian curve that best
fits the nightly averages of the data (solid line; $K_{\rm
opt}=21.7\pm1.6{\rm\,km\,s^{-1}}$), as well as the curve expected if
the neutron star has a mass of $1.4\,M_{\odot}$ (dotted line; $K_{\rm
opt}=17.5{\rm\,km\,s^{-1}}$).  The residuals to the best-fit are shown
in the middle panel.  For clarity, the error bars have been omitted.
Points taken within one night are connected with lines.  In the bottom
panel, the residuals averaged in 9 phase bins are shown.  The
horizontal error bars indicate the size of the phase bins, and the
vertical ones the error in the mean.  The dotted line indicates the
residuals expected for a $1.4\,M_{\odot}$ neutron star.}
\label{fig:vela}
\end{figure}

We have obtained about 150 new spectra, taken in as many nights, of
the optical counterpart, a B supergiant, in order to improve the mass
determination (Barziv et al.\ \cite{barz&a:01}).  These cover more
than 20 orbits, and make it possible to average out the velocity
excursions.  Unfortunately, however, we found that the average
velocity curve shows systematic effects with orbital phase (see
Fig.~\ref{fig:vela}), which dominate our final uncertainty.  While our
best estimate still gives a high mass, of $1.86\,M_\odot$, the
$2\sigma$ uncertainty of $0.33\,M_\odot$ does not allow us to exclude
soft equations of state conclusively.

While we cannot draw a firm conclusion, it is worth wondering how Vela
X-1 could be the only neutron star with a mass so different from all
others.  Barziv et al.\ (\cite{barz&a:01}) discuss this in some detail
and warned against taking the narrow mass range around $1.4\,M_\odot$
as evidence for an upper mass limit set by the EOS.  After all, for
all EOS, neutron stars substantially {\em less} massive than
$1.4\,M_\odot$ can exist, yet none are known.  Could it be that the
narrow range in mass simply reflects the formation mechanism, i.e.,
the physics of supernova explosions and the evolution of stars massive
enough to reach core collapse?  There certainly is precedent: white
dwarfs are formed with masses mostly within a very narrow range
around~$0.6~M_\odot$, well below their maximum (Chandrasekhar) mass.

Interestingly, from evolutionary calculations, Timmes et al.\
(\cite{timmww:96}) expect that single stars produce neutron stars with
a bimodal mass distribution, with peaks at 1.27 and $1.76~M_{\odot}$.
For stars in binaries, they found only a single peak at
$\sim\!1.3\,M_\odot$, but at present it is not clear whether this
result will hold (Woosley 2000, private communication).  If not, could
it be that the progenitor of Vela X-1 was a star that managed to
produce a massive neutron star?  If so, one may still wonder why no
massive radio pulsars or pulsar companions have been found.  Barziv et
al.\ (\cite{barz&a:01}) noted that this may be a selection effect: all
neutron stars with accurate masses are in binary neutron stars systems
in close orbits, whose formation requires a common-envelope stage.
During this stage, a merger can only be avoided if the initial orbit
was very wide.  Stars massive enough to form a massive neutron star,
however, likely do not evolve through a red-giant phase, and a
common-envelope phase would occur only for rather close orbits, for
which the binary would merge.

Finally, in considering the present mass measurements, one should
realise that for all neutron stars with good masses, it is expected
that they accreted only little mass after their formation.  Only
neutron stars in low-mass X-ray binaries and radio pulsars with
low-mass white dwarf companions are expected to have accreted
substantial amounts of material.  For low-mass X-ray binaries, higher
masses, of $\sim\!2\,M_\odot$, have indeed been suggested; see, e.g.,
Orosz \& Kuulkers (\cite{orosk:99}) for an analysis of Cyg~X-2, and
Zhang et al.\ (\cite{zhanss:97}) for inferences based on
quasi-periodic oscillations.  These estimates, however, rely to
greater or lesser extent on unproven assumptions.  Furthermore, for
the putative descendants, radio pulsars with white dwarf companions,
there is no evidence for such high masses (Thorsett \& Chakrabarty
\cite{thorc:99} and references therein; Van Straten et al.\
\cite{vstr&a:01}).

\section{Neutron star atmospheres}

Spectroscopic measurements of absorption lines arising in the
photosphere of a neutron star offer, in principle, one of the best
possible ways to constrain the EOS, since one could measure both the
gravitational redshift and pressure broadening, which go as $M/R$ and
$M/R^2$, respectively (Paerels \cite{paer:97}).  With the X-ray
spectrographs on board {\em Chandra} and {\em XMM}, these measurements
have become possible.  

For spectroscopy to produce quantitative results, the following is
required: (i) a source needs to have a thermal spectrum; (ii) the
spectrum should show absorption lines; (iii) pressure broadening and
pressure-induced wavelength shifts must be understood; (iv) it has to
be possible to recognise and model sources of possible additional
broadening and wavelength shifts (e.g., magnetic field).  I discuss
these points in turn.

\subsection{Thermal, photospheric emission}

For most neutron stars known, the X-ray emission is contaminated, if
not dominated, by accretion or magnetospheric processes, which are
poorly understood.  Therefore, these are unsuitable targets.

The isolated, radio-silent neutron stars, however, appear to offer a
good chance for measuring thermal, absorption spectra.  Since the
serendipitous discovery of the first of these in \cite{waltwn:96} by
Walter, Wolk, \& Neuh\"auser, five more have been uncovered in the
{\em ROSAT} All-Sky Survey (see the review by Treves et al.\
\cite{trev&a:00}).  For two sources, optical counterparts have been
identified (Walter \& Matthews \cite{waltm:97}; Motch \& Haberl
\cite{motch:98}; Kulkarni \& Van Kerkwijk \cite{kulkvk:98}).  The high
X-ray to optical flux ratios leave no model but an isolated neutron
star; stringent optical limits indicate the same for the other four
sources.

At present, it is not clear why these neutron stars are hot enough to
emit X rays, with slow accretion from the interstellar medium,
residual heat, and decay of strong magnetic fields all being
considered (e.g., Kulkarni \& Van Kerkwijk \cite{kulkvk:98}).  Most
important for the present purposes, however, is that all six sources
appear to have spectra that, as far as one can tell from current
observations, are entirely thermal.  These sources, therefore, offer
the best hope of spectra clean enough to have a chance of modeling
them reliably.

Perhaps the best-suited target for detailed study is RX
J1856.5$-$3754, the brightest steady, thermally emitting neutron star
on the sky (Walter et al.\ \cite{waltwn:96}; Walter \& Matthews
\cite{waltm:97}).  Measurements over the optical to X-ray range
indicate a spectral energy distribution very close to that of a black
body, with $kT_\infty\simeq50\,$eV (Pons et al.\ \cite{pons&a:01}).
There are some deviations, however, as indicated by, e.g., the
slightly higher best-fit black-body temperature of 63\,eV at X-ray
energies (Burwitz et al.\ \cite{burw&a:01}).  No non-thermal emission
whatsoever seems to be present, given the lack of hard X-ray emission
(Pons et al.\ \cite{pons&a:01}) and the remarkable extent to which the
optical emission is described by a Rayleigh-Jeans tail (Van Kerkwijk
\& Kulkarni \cite{vkerk:01a}).

A possible puzzle is posed by the parallax of $16.5\pm2.3\,$mas
measured by Walter (\cite{walt:01}), which, combined with the inferred
temperature of $\sim\!50\,$eV and the observed flux, implies a radius
of only $7(d/60{\rm\,pc})\,$km (Pons et al.\ \cite{pons&a:01}).  This
radius is impossible for {\em any} EOS.  Pons et al.\ suggest the
neutron star has a non-uniform temperature distribution.  For the
analysis of spectral lines, this may not a problem; more important is
the absence of non-thermal emission.

\subsection{Presence of absorption lines}

Given clean, thermal emission, the next worry is whether the spectrum
will not be too clean, i.e., whether it will show any lines.  In the
X-ray spectral range, this requires the presence of elements other
than Hydrogen and Helium in the photosphere (unless very strong
magnetic fields are present).

In general, one would expect that, if any Hydrogen is present,
gravitational settling would make it float on top.  This process is
fast and thus a pure Hydrogen atmosphere might form, resulting in a
line-less X-ray spectrum.  Observationally, the situation looks
promising: the spectral energy distributions of the two sources with
optical counterparts are similar to those of black bodies.  This is
{\em in\/}consistent with pure (unmagnetised) hydrogen or helium
atmospheres, whose virtual transparency at X-ray wavelengths would
lead -- just as in, e.g., Sirius B -- to strongly non-black-body
spectral shapes (Pavlov et al.\ \cite{pavl&a:96}).  Since models with
solar abundance, pure iron, or `Si-ash' abundances all predict strong
absorption features (Romani \cite{roma:87}; Rajagopal \& Romani
\cite{rajar:96}; Zavlin, Pavlov, \& Shibanov \cite{zavlps:96}; Pons et
al.\ \cite{pons&a:01}; G\"ansicke, Braje, \& Romani \cite{gaenbr:01}),
the prospects looked good.

Reality, however, was disappointing the first spectrum at good
resolution, taken with {\em XMM} of RX~J0720.4$-$3125 showed no
features (Paerels et al. \cite{paer&a:01}).  Furthermore, the spectra
of the Vela pulsar and of PSR~B0656+14, both of which have relatively
strong thermal components, appeared to be completely featureless as
well (Pavlov et al.\ \cite{pavl&a:01}; H. Marshall, priv.\ comm.).
And finally, also for RX J1856.5$-$3754, X-ray spectra taken with {\em
Chandra} (Burwitz et al.\ \cite{burw&a:01}), ultra-violet spectra
taken with {\em HST} (Pons et al.\ \cite{pons&a:01}), as well as
optical spectra taken with VLT (Van Kerkwijk \& Kulkarni
\cite{vkerk:01a}) failed to show strong features.

\begin{figure*}
\centerline{\epsfig{figure=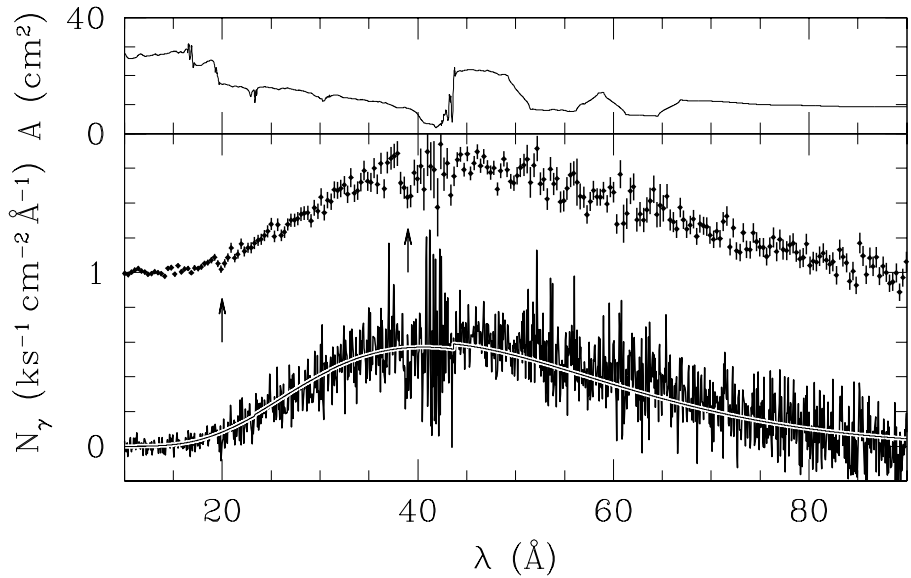,width=0.9\textwidth}}
\caption[]{Presently available 56\,ks LETG+HRC-S spectrum of RX
J1856.5$-$3754, calibrated using effective areas determined at
SRON/Utrecht (top panel).  The top curve is binned to 0.34\,\AA, the
bottom one to 0.06\,\AA.  Overdrawn is the best-fit black-body as
derived by Burwitz et al.\ (\cite{burw&a:01}): $kT_\infty=63\pm3\,$eV,
$N_{\rm H}=(1.03\pm0.20)\times10^{20}{\rm\,cm^{-2}}$,
$R_\infty/d=0.037\pm0.005{\rm\,km\,pc^{-1}}$.  Unlike other neutron
stars observed so far, there are absorption features!  These are
marked by arrows.  The one at 39\,\AA\ appears highly significant,
while that at 20\,\AA\ seems marginal.  The longer integration
currently planned in director's discretionary time will allow to
confirm (or not) their reality, as well as to detect other lines.  If
multiple lines are present, this spectrum may well provide the best
constraint yet on the EOS.\label{fig:rxj1856}}
\end{figure*}

The situation may not be completely hopeless, however, as our own
analysis of the LETG data did turn up possible features, at $\sim\!39$
and 20\,\AA\ (Fig.~\ref{fig:rxj1856}).  At present, it is not clear
what these could be due to.  As mentioned, pure Hydrogen and Helium
atmospheres are excluded by the overall spectral shape, and pure
heavy-element atmospheres by the lack of much stronger features.  We
have considered whether low-level accretion might lead to metals in
the photosphere.  This can be estimated using the density of the
interstellar medium inferred from the cometary H$\alpha$ nebula around
the source that we discovered with the VLT (Van Kerkwijk \& Kulkarni
\cite{vkerk:01b}), under the assumption that this nebula is due to
ionisation (see below for an alternative interpretation).  We infer an
accretion rate of about $10^9{\rm\,g\,s^{-1}}$.  This is far less than
required to power the X-ray luminosity of the source, and the
photospheric metal abundances will be very low, due to rapid settling.
Even at fractional abundances by number of $10^{-7}\ldots10^{-6}$,
however, it appears lines might still form, although it is unclear
whether the continuum shape could be made to fit the observations.

In considering the above, one should realise that the current
integrations have not been very long, and the signal-to-noise ratios
rather low (see Fig.~\ref{fig:rxj1856}).  Indeed, at the current
signal-to-noise and resolution, it would be hard to classify an
optical spectrum of a star like the Sun.  In AO3, therefore, there
were a number of proposals to obtain a longer {\em Chandra}
observation on RX J1856.5$-$3754.  None of these made it, but the
director, H. Tananbaum, has just announced that he will use his
discretionary time to observe the source for 450\,ks.  Hence, we
should soon know more!

\subsection{Model atmospheres}

From absorption lines such as those seen in Fig.~\ref{fig:rxj1856}, it
is straightforward to measure centroids accurate enough to determine a
good gravitational redshift ($z_{\rm GR}\simeq0.3$).  Furthermore, the
lines are well resolved, and thus it should be possible to measure the
pressure broadening and therewith the surface gravity.

First of all, however, one has to identify the lines.  With the two
putative lines in the spectrum of RX J1856.5$-$3754, this is not yet
possible.  Even with identifications, the interpretation may pose a
problem, as our physical understanding may not suffice to make
reliable predictions of pressure broadening and pressure-induced
wavelength shifts.  For instance, current treatment of overlapping
Stark-broadened excitation levels (e.g., in the Opacity Project;
Seaton et al.\ \cite{seatmp:94}) is {\em known to fail\/} even under
the relatively low pressures encountered in white dwarfs (Bergeron
\cite{berg:93}), leading to bad fits to Hydrogen Balmer and especially
Lyman profiles (Finley et al.\ \cite{finl&a:97}), both for white-dwarf
spectra and for spectra taken of high-pressure plasmas on Earth.
Clearly, the fact that, as yet, we lack sufficient understanding of
even Hydrogen, should lead to some moderation in ones claims.  The
problems do not appear to be of fundamental nature, however, and with
{\em observed} features in neutron-star atmospheres, there should be
ample incentive for further study.

\subsection{Other physical effects}

Strong magnetic fields ($h\nu_{\rm cyc}>(kT,h\nu_\gamma)\Rightarrow
B\ga10^9\,$G at X-ray wavelengths), as are found in radio pulsars,
could cause additional broadening and wavelength shifts.  For RX
J0720.4$-$3125 and RX J0420.0$-$5022, 8-s and 23-s periodicities
(Haberl et al.\ \cite{habe&a:97}, \cite{habe&a:99}) provide indirect
evidence for a magnetic field.  Indeed, it may well be that these
sources are middle-aged `magnetars,' with
$B\simeq10^{14}\ldots10^{15}\,$G, which are kept hot by magnetic field
decay (Kulkarni \& Van Kerkwijk \cite{kulkvk:98}; Heyl \& Kulkarni
\cite{heylk:98}).

For RX~J1856.5$-$3754, very strong fields seem to be excluded by the
absence of pulsations (Pons et al.\ \cite{pons&a:01}; Burwitz et al.\
\cite{burw&a:01}).  A weaker field might nevertheless be present.
Indeed, an alternative interpretation for the H$\alpha$ nebula around
RX~J1856.5$-$3754 is that it is a bow-shock nebula, such as seen
around a number of pulsars (PSR~B1957+20: Kulkarni \& Hester
\cite{kulkh:88}; PSR B2224+65: Cordes, Romani, \& Lundgren
\cite{cordrl:93}; PSR J0437$-$4715: Bell et al.\ \cite{bell&a:95}).
If so, a relativistic wind must be present, which presumably arises in
a magnetosphere.  In order to avoid pulsations as well as any sign of
non-thermal emission at optical and X-ray wavelengths, one might
appeal to a rather weak, few $10^{11}\,$G field (Van Kerkwijk \&
Kulkarni \cite{vkerk:01b}).  The question remains, however, whether
the presence of such field, or a magnetic field in general, could lead
to spectra as close to those of black bodies as observed.

A magnetic field might also induce absorption features.  For instance,
the ratio of roughly a factor two in wavelength between the two
possible features we observe, led G.~Pavlov (2000, priv.\ comm.) to
speculate that they might represent proton and $A/Z=2$ cyclotron lines
in a strong, $\sim\!10^{14}\,$G field.  If confirmed, the spectra will
likely not be very useful to constrain the EOS, but it would
constitute the first direct proof of an ultra-strong magnetic field in
a neutron star.

\section{Prospects}

In the near future, it should be possible to increase the number of
accurate neutron-star mass determinations substantially.  For X-ray
pulsars other than Vela X-1, it is straightforward to obtain better
radial-velocity amplitudes.  By choosing binaries with circular
orbits, the problems that beset the mass determination for Vela X-1
can largely be avoided.  For radio pulsars, of most interest seem to
be those which should have accreted a substantial amount of matter,
i.e., the millisecond pulsars.  With only little improvement, optical
studies of their companions such as those done by Van Kerkwijk,
Bergeron, \& Kulkarni (\cite{vkerbk:96}) and Callanan, Garnavich, \&
Koester (\cite{callgk:98}), and precise radio timing studies such as
the marvellous work by Van Straten et al.\ (\cite{vstr&a:01}) can lead
to accurate masses.

X-ray spectroscopy of neutron stars has just started, and has led to
both disappointments and hope.  The longer observations with {\em XMM}
and {\em Chandra} currently planned, in particular the 450\,ks LETG
observation in director's discretionary time of RX~J1856.5$-$3754,
will certainly help to learn about neutron-star atmospheres.  What
exactly we will learn one cannot say, since at present we seem to lack
even basic understanding, being unable to answer as (apparently)
simple a question as why the spectra are so close to black bodies.
Given that, it is also unclear whether this avenue can lead to useful
constraints the EOS, but at least in principle the prospects are very
good.  Certainly, exciting times are ahead.

\end{document}